# Effect of Ultrahigh Stiffness of Defective Graphene from Atomistic Point of View

D.G. Kvashnin,[†,‡,*] P.B. Sorokin [†,‡,§,⊥]

[†] National University of Science and Technology MISiS, 4 Leninskiy prospekt, Moscow, 119049, Russian Federation

[‡] Emanuel Institute of Biochemical Physics RAS, 4 Kosigina Street, Moscow, 119334, Russian Federation

[§] Moscow Institute of Physics and Technology, 9 Institutsky lane, Dolgoprudny, 141700, Russian Federation;

[⊥] Technological Institute for Superhard and Novel Carbon Materials, 7a Centralnaya Street, Troitsk, Moscow, 142190, Russian Federation

*dgkvashnin@gmail.com

Well-known effect of mechanical stiffness degradation under the influence of point defects in macroscopic solids can be controversially reversed in the case of low-dimensional materials. Using atomistic simulation, we showed here that a single-layered graphene film can be sufficiently stiffened by monovacancy defects at a tiny concentration. Our results correspond well with recent experimental data and suggest that the effect of mechanical stiffness augmentation is mainly originated from specific bonds distribution in the surrounded monovacancy defects regions. We showed that such unusual mechanical response is the feature of presence of specifically monovacancies, whereas other types of point defects such as divacancy, 555-777 and Stone-Wales defects, lead to the ordinary degradation of the graphene mechanical stiffness.

Isolation of the single-layered graphene films immediately settled a novel rapidly growing area in the condensed matter physics. Additionally to its impressive electronic properties, mechanical stiffness of graphene plays one of the most important roles for further technological applications. Two-dimensional nature of graphene combined with stiff covalent bonding between carbon atoms lead to the high elasticity and enormous stiffness of the film.[1] Such features of the mechanical properties suggest a wide range of possible applications including armor,[2] high-strength device elements,[3] composites reinforcing,[4] or protection coating.[5]

Typically, mechanical properties of the 2D structures are measured using the nanoindentation technique when films are probed by indenting with a tip of an atomic force microscope. Such test can be treated as a local measurement of strength, whereas in-plane tests measure the global strength of the films. Indentation technique widely used for investigation of mechanical characteristics of surfaces,[6] and nanostructures such as carbon nanotubes,[7] graphene[1], and MoS$_2$.[8] Nevertheless, the measured in-plane elastic stiffness constants in both approaches are close to each other and display the highest value of 1.1 TPa[1] which is close to graphite. However, mechanical response of the 2D films fundamentally differs from the bulk counterpart. Data in Ref. 9 demonstrated that local strength of a region with 1D defects (grain boundaries, dislocations, disclinations) in graphene is higher than that of the regions with a perfect structure. It suggests the specific impact of structural defects on the mechanical properties of graphene. The vast majority of previous studies on the mechanical response of graphene that contained point defects [10-14] supposed the ordinary degradation of the film stiffness and only recently our suggestion [15] that graphene can be significantly stiffened by inclusion of small number (concentration <1%) of point vacancy defects was fully confirmed by the experimental report.[16] Here we investigate this controversial idea that the inclusion of point defects in graphene leads to a *strong increase* in the film stiffness.

We study this effect in detail by using comprehensive simulation of defective graphene. We applied a well-developed theoretical approach of Brenner bond-order potential [17] to directly simulate the process of graphene indentation. In order to validate the obtained results for the graphene with monovacancy defects, we additionally used the Tersoff many-body potential.[18] All calculations were performed using a LAMMPS simulation package.[19] We simulated graphene of a circular shape (Figure 1a) with fixed boundaries indented by a spherical geometry body with the mean diameter equal to 1/10 of mean diameter of the films. This simulation setup was chosen according to experimental data where such process was carried out.[1,16,20] Interaction between the film and the simulated tip was described with a pure repulsive force. Deflection was carried out with a step of 1.2 Å. Before the mechanical test and at each indentation step, the system was relaxed using conjugated gradient minimization while the maximum interatomic forces became less or equal to 0.05 eV/Å. To exclude the effect of stiffness constant dependence on the relative position of the tip,[9] we performed 20 computational tests for each concentration value with random defects distribution. The defects in the system were created by removing atom (in the case of vacancy defects) or bond rotation (in the case of Stone-Wales defects). Also reconstructed 555-777 defect was considered (divacancy with rotated bond). The connection of defects and their allocation on the fixed boundary was not allowed.

Graphene mechanical response to indentation is represented by increasing of indentation force with deflection of the film atoms directly under the indenter. From this relation, the local fracture force of the film by the indentation force at fracture can be calculated, whereas the whole nonlinear force-deflection dependence can be used for calculation of the stiffness constant $E^{2D}$ using the following relation. [1,16,20]

$$F(\delta) = \sigma^{2D}(\pi a)\left(\frac{\delta}{a}\right) + E^{2D}\left(q^3 a\right)\left(\frac{\delta^3}{a^3}\right). \tag{1}$$

Here $\delta$ is the deflection of the atom in the center point, $a$ is the film radius, $q = \dfrac{1}{1.05 - 0.15\nu - 0.16\nu^2}$ is a dimensionless constant dependent from the Poisson ratio, [1] $\sigma^{2D}$ is the pretension in the film. The linear term in the equation plays a major role at the small load, whereas the second term dominates for large deflection. From Eq. (1) Young's modulus can be obtained by dividing of $E^{2D}$ value to the formal thickness of graphene sheet defined as 3.34 Å (van der Waals distance between layers in graphite [1]).

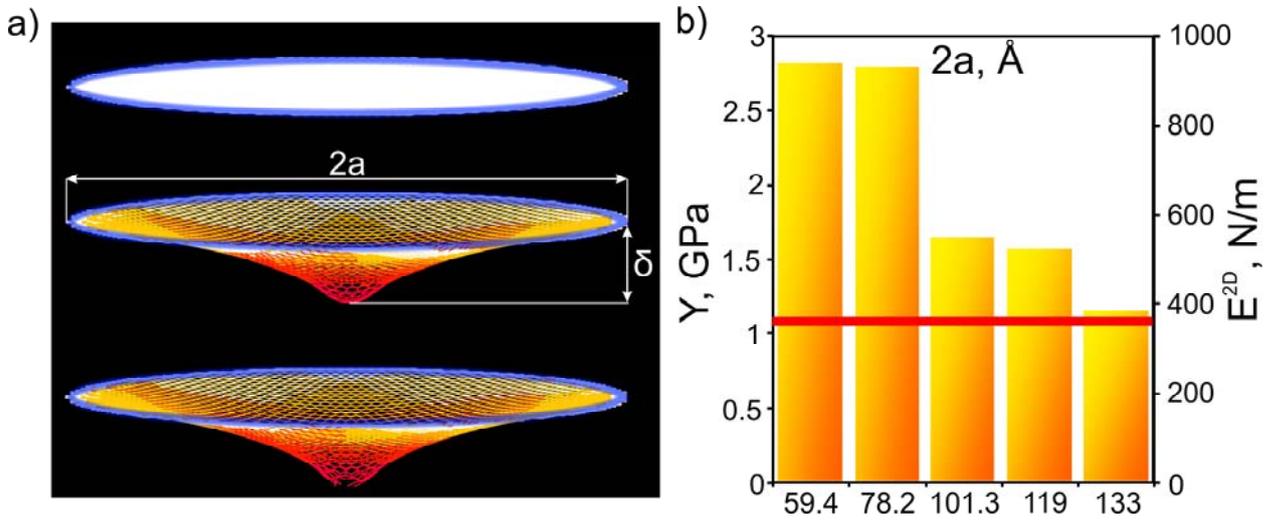

**Figure 1. a) The process of graphene deformation including the initial, critically strained, and fractured structure. The color variation represents the bond lengths, from white (1.39-1.42 Å) to red (1.46-1.54 Å). Bonds of the fixed boundary atoms are marked by blue; b) Dependence of graphene stiffness constant value on the film diameter. Alternated vertical axes correspond to 3D ($Y$) and 2D ($E^{2D}$) values. Experimental result [1] is marked by a horizontal red line.**

At the first stage, we calculated the stiffness constant of perfect graphene with various diameters (from 5.9 to 13.3 nm). All films display similar behavior of bonds elongation with the deflection (Figure 1a) as well as force-deflection curves. A monotonically decreasing $E^{2D}$ value with a tendency to the experimental value [1] was obtained (Figure 1b). We found that graphene with a 13.3 nm diameter is large enough to be free from the influence of the boundaries and displays a 381.6 N/m (1.13 TPa) stiffness constant close to the experimental one. In further work, graphene with this diameter was used.

Next, we simulated the behavior of graphene sheets with monovacancies concentration in the range from 0 to 1.5% and observed the specific behavior of the force-deflection curves. Though the critical deflection depth is normally lower than that for perfect graphene (membrane became more brittle), the fracture force displays a higher value. The fracture force became higher than that of perfect graphene from the 0.1% monovacancy defect concentration and reached a maximal value at the 0.2% concentration (125.4 nN vs. 102.2 nN for the perfect graphene), see Figure 2a. It unambiguously states that graphene becomes stiffer. We calculated the stiffness constant and found that at the range of defect concentration from 0.1% to 0.6%, a significant increase in the stiffness of graphene occurs (Figure 2b). The obtained data closely correspond with the reference experiment [16] (Figure 2b-inset), which allows to suppose that during the $Ar^+$ irradiation mostly monovacancy defects were formed. The results of atomistic simulations by the Brenner potential were validated by the Tersoff potential (a blue curve) which showed a similar increasing of $E^{2D}$ for the defect concentration from 0.1 to 0.6%. The error distribution in the data calculated by the Brenner and Tersoff approaches varied from ±22.8 N/m (±68.1 GPa) to ±49.3 N/m (±147.2 GPa) and from ±40.6 N/m (±121.2 GPa) to ±111.9 N/m (±334.1 GPa), respectively (see also Supplementary materials Figure S1).

Further increasing of the defect concentration leads to a decrease in the stiffness constant, the calculated values of $E^{2D}$ for 0%, 1% and 1.5% monovacancy concentrations display an almost linear dependence which perfectly corresponds with the previous theoretical data where similar degradation of the graphene stiffness with a rising defect concentration was reported. [13]

It should be noted that such special dependence of the elastic constant obtained by nanoindentation on the defect concentration does not take place in the case of the elastic constant obtained by in-plane deformation of the film (Figure 2b). [21] The latter value was calculated using the equation $C = \frac{1}{A} \frac{\partial^2 E}{\partial \varepsilon^2}$ (where $A$ is area of the 2D graphene unit cell, $E$ is the strain energy, $\varepsilon$ is the in-plane strain in armchair direction. The partial derivatives at zero strain in other dimensions are vanished which yields $C$ as an analog of the elastic constant $C_{11}$ of bulk graphite). The in-plane stiffness linearly decreases with increasing of the vacancy concentration. The slope of the dependence ($15.95 \cdot 10^{-12}$ N/m·cm$^2$) corresponds very well with the reference result of tight binding calculation ($15.81 \cdot 10^{-12}$ N/m·cm$^2$) [21] which additionally validates used approach. This result allows for the conclusion that the effect of stiffening is connected with a specific kind of deformation by the probe indentation.

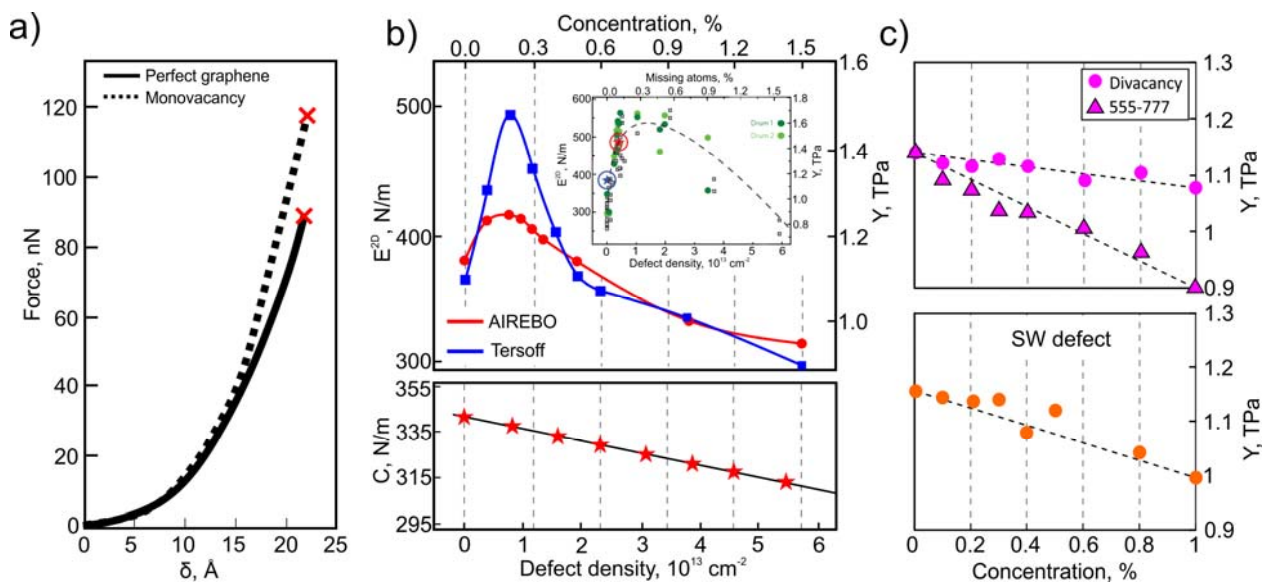

**Figure 2.** a) indentation force-deflection curves for perfect graphene and graphene containing 0.2% vacancy defects; fracture forces are indicated by crosses; b) top section: the dependence of graphene stiffness constant value on monovacancy concentration calculated using the Brenner (red curve) and Tersoff (blue curve) empirical potentials. Alternated vertical axes correspond to 2D and 3D values. In the inset, the experimental data from Ref. 16 is shown; bottom section: the dependence of in-plane stiffness on monovacancy concentration; c) the behavior of stiffness constant of graphene containing both divacancies and 555-777 (pink circles and pink triangles, respectively) and Stone-Wales (orange dots) defects. To avoid data overloading, only average values are presented in the calculated stiffness constant dependences, see Supplementary materials Figure S1 (http://pubs.acs.org/doi/suppl/10.1021/acs.jpclett.5b00740) for the dependences with depicted errors.

The effect of graphene stiffening strongly depends on the point defect type. For example, introducing of divacancy or Stone-Wales (SW) defects in the structure leads to the practically vanishing of the augmentation effect (see Figure 2c). We considered unreconstructed and reconstructed (r-divacancy or 555-777 defects) divacancy defects. Both of them were observed in graphene.[22]

The nature of augmentation of the stiffness constant was investigated by direct analysis of the surrounded defects regions in deflection just before the fracture. The energy and bond length distributions were calculated for all considered films with mono-, divacancies, 555-777 and Stone-Wales point defects. In Figure 3a distribution of energy per atom for a 0.2% defect concentration of all mentioned types at a critical value of deflection is shown. From the per-atom energy distribution (Figure 3a) it can be seen that the in the case of graphene containing mono- and divacancy defects, two coordinated atoms located directly on the defects accumulated the energy more than two times higher (yellow color) and undertook a much higher loading than the atoms in the rest of graphene area (blue color). In the case of defects with rotated bonds (555-777 and SW defects), such energy distribution was not observed.

Additionally, we found out different behavior of the bonds between the atoms located directly on the defects and bonds in the graphene area under loading. In the case of graphene with monovacancy defects, the bonds between two coordinated carbon atoms stretch much lower. At critical deflection, the difference between the bond lengths of two and three coordinated atoms reaches 2%. In the case of graphene with divacancy defects, this effect is two times weaker, whereas graphene containing 555-777 and Stone-Wales defects displays the opposite behavior of the bond lengths: the bonds between the atoms in the defect area are longer than the bonds in the graphene area.

These results allow for the conclusion that a missing atom leads to hardening of the nearby area: bond length becomes shorter and the bonds can undertake much more loading than those in the remaining graphene area. Due to the purely local character of the observed effect, it depends on the defects concentration and size: although the similar behavior of the bonds was observed in both mono- and divacancy cases, only the former can lead to graphene stiffening, because as the defect size increases, the graphene lattice becomes too sparse and weak.

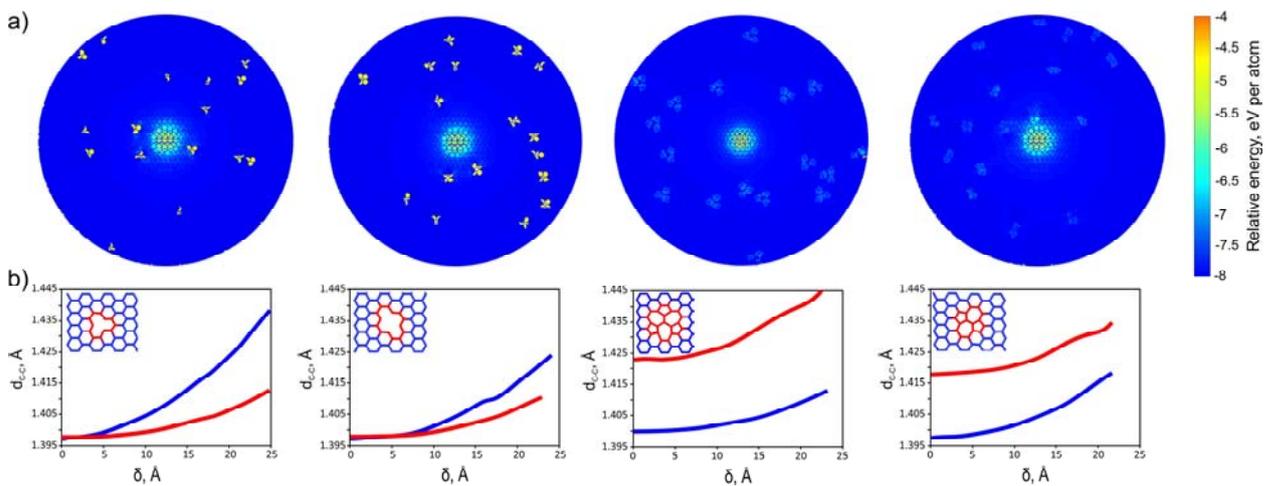

Figure 3. The energy distribution and bond lengths behavior of graphene with the highest value of stiffness constant $E^{2D}$ = 451.01 N/m (monovacancy defect concentration 0.2%). a) Energy per atom distribution of mono-, divacancies, 555-777 and Stone-Wales defect concentration at a critical value of deflection; b) the dependence of bond length between the atoms located directly on the defects (excluding defects directly under the tip) and bonds in the remaining graphene area under the film loading.

The significant augmentation of graphene stiffness clearly manifests that graphene not only holds the place of the stiffest material but can be further hardened in a specific way. The pure 2D nature of this effect demonstrates the importance of further developing of the monoatomic films elastic theory taking into account the small atomistic effects which can fundamentally change the properties of the whole material.

ACKNOWLEDGEMENTS. We thank Nature Physics for the kind permission to reproduce results (further material and references are given in the figure 2 caption): Fig. 3a Nature Physics 11, 26 (2015) (G. López-Polín et al.). This work was supported by the Grant of the President of the Russian Federation for government support of young PhD scientists MK-6218.2015.2 (project ID 14.Z56.15.6218-MK). Authors acknowledge Prof. Leonid A. Chernozatonskii and Prof. David Tománek for fruitful discussions. We are grateful to the 'Chebyshev' and 'Lomonosov' supercomputers of the Moscow State University for the possibility of using a cluster computer for our simulations. Part of the calculations was made on the Joint Supercomputer Center of the Russian Academy of Sciences.